\documentclass[11pt]{cernrep}

\usepackage{graphicx}

\usepackage{cite,./mcite}

\newcommand{\bs}{\begin{slide}}
\newcommand{\es}{\end{slide}}
\newcommand{\be}{\begin{equation}}
\newcommand{\ee}{\end{equation}}
\newcommand{\bea}{\begin{eqnarray}}
\newcommand{\eea}{\end{eqnarray}}

\newcommand{\la}{\left\langle}
\newcommand{\ra}{\right\rangle}
\newcommand{\lc}{\left[}
\newcommand{\rc}{\right]}
\newcommand{\lp}{\left(}
\newcommand{\rp}{\right)}

\newcommand{\aq}{\alpha_s\lp Q^2\rp}
\newcommand{\aqq}{\alpha_s\lp Q_0^2\rp}
\newcommand{\bc}{\begin{center}}
\newcommand{\ec}{\end{center}}

\newcommand{\bi}{\begin{itemize}}
\newcommand{\ei}{\end{itemize}}
\def\epm#1#2{\hbox{${\lower1pt\hbox{$\scriptstyle +~#1$}}
\atop {\raise1pt\hbox{$\scriptstyle -~#2$}}$}}
\newcommand{\dat}{\mathrm{dat}}
\newcommand{\art}{\mathrm{art}}
\newcommand{\net}{\mathrm{net}}
\newcommand{\rep}{\mathrm{rep}}
\newcommand{\rmexp}{\mathrm{exp}}

\begin{document}

\title{The neural network approach to parton distributions \\
HERA - LHC Workshop Proceedings }

\author{L. Del Debbio$~^a~$, S. Forte$~^b~$, 
J.I. Latorre$~^c~$, A. Piccione$~^d~$ and
J. Rojo$~^c~$ (The NNPDF Collaboration) }

\institute{ $~^a$Theory Division, CERN, $~^b$ Dipartimento di Fisica, 
Universit\`a di Milano and 
INFN, Sezione di Milano, $~^c$ Departament d'Estructura i 
Constituents de la Mat\`eria, 
Universitat de Barcelona and \\$~^d$ Dipartimento di Fisica Teorica, 
Universit\`a di Torino, and
INFN Sezione di Torino.  }

\maketitle

\begin{abstract}
We introduce the neural network approach to global fits
of parton distribution functions. First we review previous
work on unbiased parametrizations of deep-inelastic 
structure functions with 
faithful estimation of their uncertainties, and then we
summarize the current status of neural network parton distribution fits.

We introduce the neural network approach to global fits
of parton distribution functions. First we review previous
work on unbiased parametrizations of deep-inelastic 
structure functions with 
faithful estimation of their uncertainties, and then we
summarize the current status of neural network parton distribution fits.

\end{abstract}



%
%

The requirements of precision physics at
 hadron colliders, as has been emphasized through
this workshop, have recently led
to a rapid improvement in the techniques for the determination of 
parton distribution functions (pdfs) of the nucleon.
Specifically it is now mandatory to 
determine accurately the uncertainty on these quantities, and
the different collaborations performing global pdf 
analysis\cite{mrst01e,*cteq61,*ale02} have performed 
estimations of these uncertainties using a variety
of techniques. The
main difficulty 
is that one is trying to determine the uncertainty on a function,
that is, a probability measure in a space of functions, and to extract it from
a finite set of experimental data, 
a problem which is mathematically ill-posed. 
It is also known that the standard approach to global parton fits 
have several shortcomings: 
the bias introduced by choosing fixed functional forms to
parametrize the parton distributions
(also known as {\it model dependence}), the problems to assess
faithfully the pdf uncertainties, the
combination of inconsistent experiments, and the lack
of general, process-independent error propagation techniques.
Although the problem of quantifying the uncertainties in pdfs has seen
a huge progress since its paramount importance was raised some
years ago, until now no unambiguous conclusions have been obtained.

In this contribution we present a novel strategy to address the problem
of constructing unbiased parametrizations of parton distributions
with a faithful estimation of their uncertainties, based on 
a combination of two techniques: Monte Carlo methods and neural networks.
This strategy, introduced in \cite{f2nn,nnpdf},
has been first implemented to address the marginally simpler problem
of parametrizing deep-inelastic structure functions $F(x,Q^2)$, which
we briefly summarize now. 
In a first step we construct a Monte Carlo sampling of the experimental data 
(generating artificial data replicas), and then
we train neural networks to each data replica, to
construct a probability measure in the space of structure functions
$\mathcal{P}\lc F(x,Q^2)\rc$. The probability measure constructed
in this way
contains all information from experimental data, including correlations,
with the only assumption of smoothness. Expectation values and moments over
this probability measure are then evaluated as averages over
the trained network sample,
\be
\label{probmeas}
\la \mathcal{F}\lc F(x,Q^2)\rc\ra=\int\mathcal{D}F
\mathcal{P}\lc F(x,Q^2)\rc
\mathcal{F}\lc F(x,Q^2)\rc=\frac{1}{N_{\rep}}
\sum_{k=1}^{N_{\rep}}\mathcal{F}\lp F^{(\net)(k)}(x,Q^2)\rp \ .
\ee
where $\mathcal{F}\lc F\rc$ is an arbitrary function of $F(x,Q^2)$.

The first step is the Monte Carlo sampling of experimental data, 
generating $N_{\rep}$ replicas of the original $N_{\dat}$ experimental data,
\be
F_i^{(\art)(k)} =\lp 1+r_N^{(k)}\sigma_N\rp\lc F_i^{(\rmexp)}+r_i^{s,(k)}
\sigma^{stat}_i+\sum_{l=1}^{N_{sys}}r^{l,(k)}\sigma_i^{sys,l} \rc, 
\qquad i=1,\ldots,N_{\dat} \ ,
\ee
where $r$ are gaussian random numbers with the same correlation
as the respective uncertainties, and $\sigma^{stat},\sigma^{sys},
\sigma_{N}$ are the statistical, systematic and normalization
errors.
The number of replicas $N_{\rep}$ has to be large enough so that the
replica sample
 reproduces central values, errors and correlations
of the experimental data.

The second step consists on training a neural network\footnote{For
a more throughly description of neural network, see \cite{f2nn}
and references therein}  on each of the data
replicas. Neural networks
are specially suitable to parametrize parton distributions since
they are unbiased, robust approximants and interpolate between
data points with the only assumption of smoothness. The neural network 
training consist on the minimization for each replica of the
$\chi^2$ defined with the inverse of the 
experimental covariance matrix,
\be
{\chi^2}^{(k)}=\frac{1}{N_{\dat}}\sum_{i,j=1}^{N_{\dat}}\lp
F_i^{(\art)(k)}-F_i^{(\net)(k)}\rp\mathrm{cov}^{-1}_{ij}
\lp F_j^{(\art)(k)}-F_j^{(\net)(k)}\rp \ .
\ee
Our minimization strategy is based on
Genetic Algorithms (introduced in \cite{rojo04}), which are specially suited
for finding global minima in highly nonlinear 
minimization problems.


The set of trained nets, once is validated through suitable statistical
estimators, becomes the sought-for probability measure 
$\mathcal{P}\lc F(x,Q^2)\rc$ in the space of structure functions. 
Now observables with
errors and correlations can be computed from averages over this
probability measure, using eq. (\ref{probmeas}). 
For example, the average and error of a
structure function $F(x,Q^2)$ at arbitrary $(x,Q^2)$ can be
computed as
\be
\la F(x,Q^2) \ra =\frac{1}{N_{\rep}}\sum_{k=1}^{N_{\rep}}
F^{(\net)(k)}(x,Q^2), \quad
\sigma(x,Q^2)=\sqrt{\la  F(x,Q^2)^2 \ra-\la F(x,Q^2) \ra^2} \ .
\ee
A more detailed account of the application
of the neural network approach to structure
functions can be found in \cite{nnpdf}, which describes
the most recent NNPDF parametrization of the proton structure 
function\footnote{
The source code, driver program and graphical web interface for
our structure function fits is available at {
\tt http://sophia.ecm.ub.es/f2neural}.}.

Hence this strategy can be used also to parametrize
parton distributions, provided one now takes into 
account perturbative QCD evolution. Therefore we need to define
a suitable evolution formalism. Since complex
neural networks are not allowed, 
we must use the convolution theorem to evolve
parton distributions in $x-$space using the inverse $\Gamma(x)$
of the Mellin space evolution factor $\Gamma(N)$, defined as
\be
q(N,Q^2)=q(N,Q_0^2) 
\Gamma\lp N,\aq,\aqq\rp \ ,
\ee
The only subtlety is that the x-space evolution
factor $\Gamma(x)$ is a distribution,
which must therefore be
regulated at $x=1$, yielding the final evolution equation,
\be
q(x,Q^2)= q(x,Q_0^2)\int_x^1 dy~\Gamma(y) +\int_x^1\frac{dy}{y}
\Gamma(y)\lp q\lp\frac{x}{y},Q_0^2\rp -yq(x,Q_0^2)\rp \ ,
\ee
where in the above equation $q(x,Q_0^2)$ is parametrized
using a neural network.
At higher orders in perturbation theory coefficient functions $C(N)$
are introduced through a modified evolution factor, $\tilde{\Gamma}(N)\equiv
 \Gamma(N) C(N)$.
We have benchmarked our evolution code with the
Les Houches benchmark tables \cite{lh} at NNLO up to an
accuracy of $10^{-5}$.
The evolution factor $\Gamma(x)$ and its integral are
computed and interpolated before
the neural network training in order to have a faster fitting
procedure.

\begin{figure}
\includegraphics[scale=0.4]{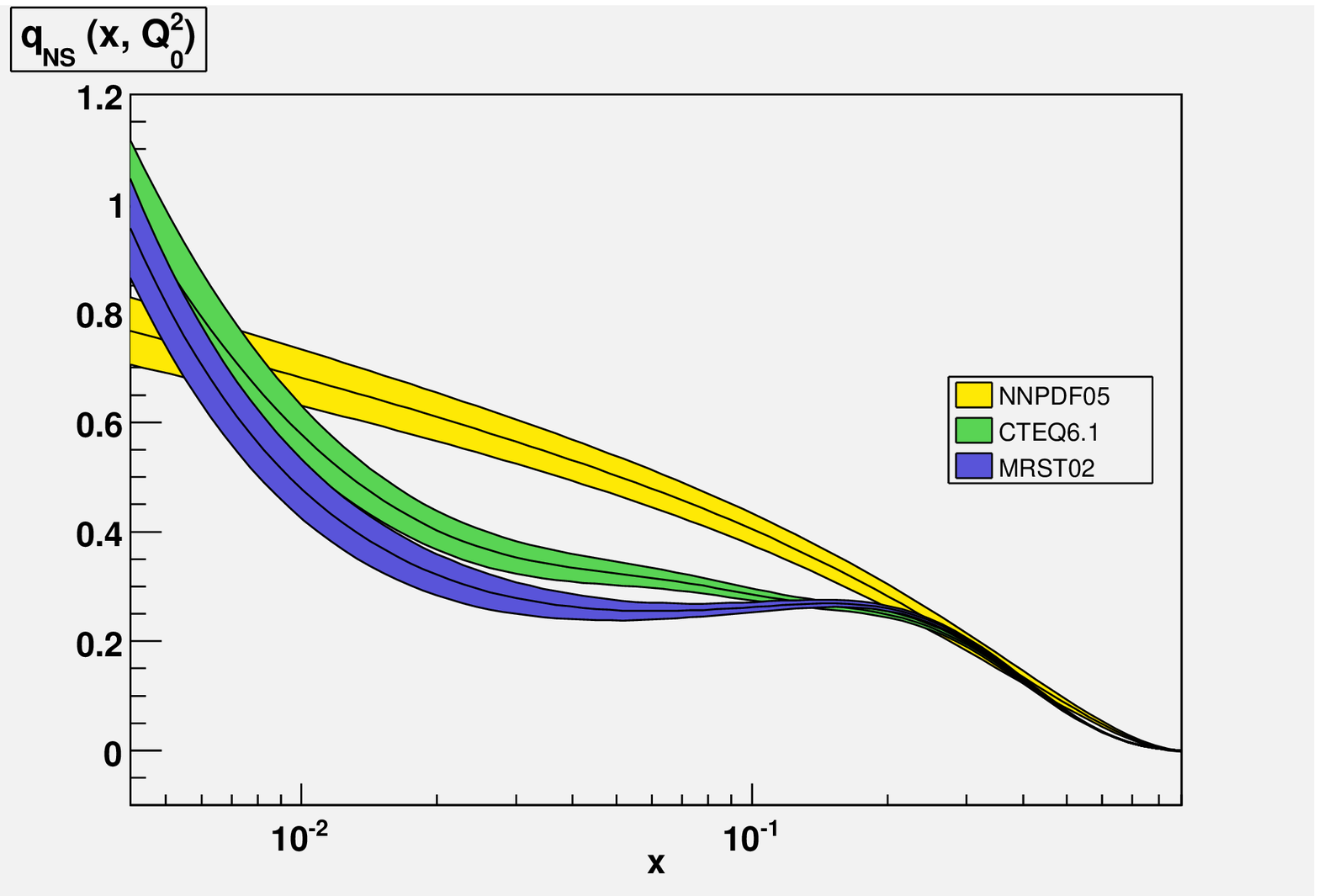}
\includegraphics[scale=0.4]{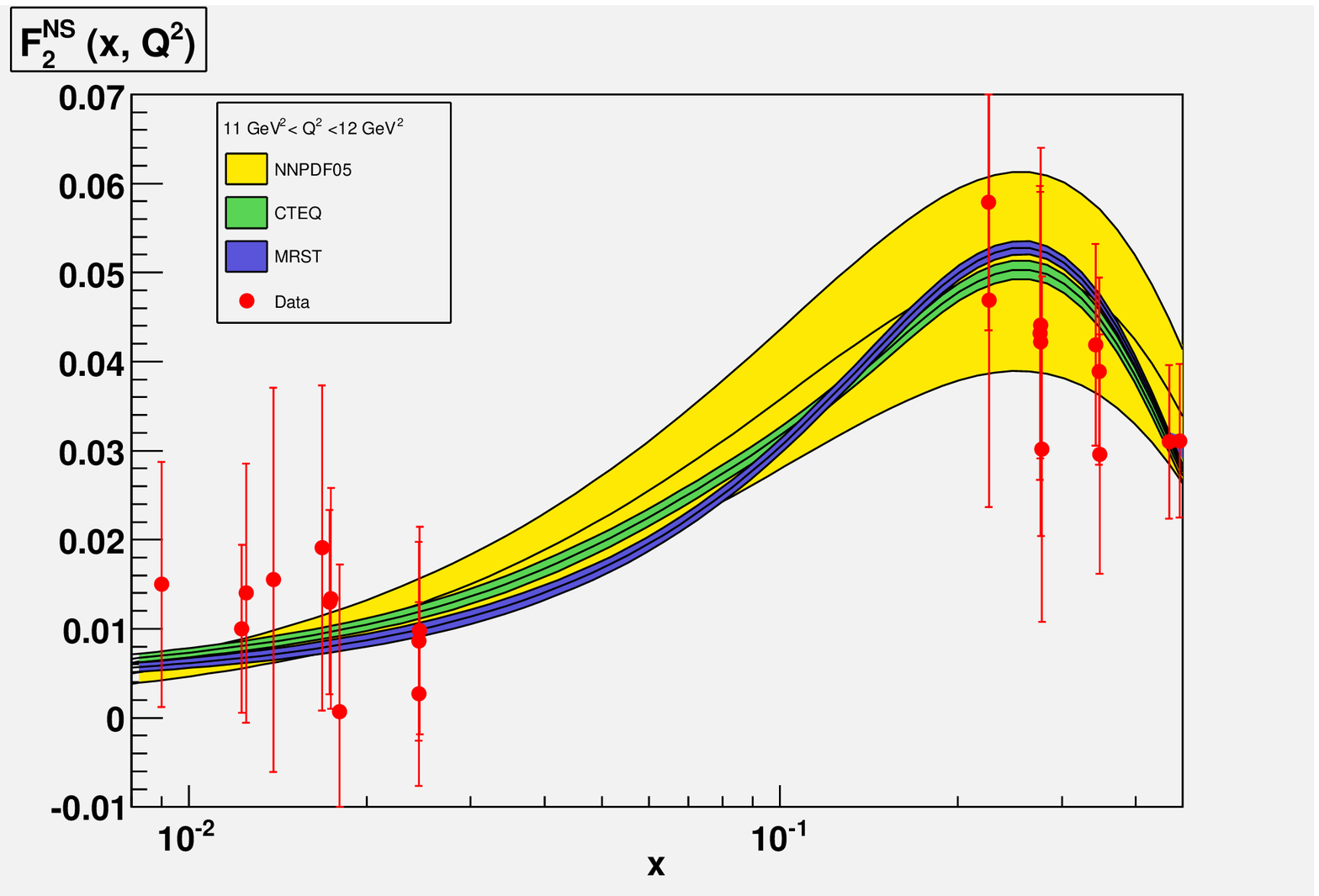}
\caption{Preliminary results for the 
NNPDF $q_{NS}$ fit at $Q_0^2=2\,\mathrm{GeV}^2$, and
the prediction for $F_2^{NS}(x,Q^2)$ compared with the CTEQ and MRST
results.}
\label{f2nn}
\end{figure}

As a first application of our method, we have extracted the nonsinglet
parton distribution $q_{NS}(x,Q^2_0)=\frac{1}{6}\lp
u+\bar{u}-d-\bar{d}\rp(x,Q^2_0)$ from the nonsinglet structure
function $F_2^{NS}(x,Q^2)$ as measured by the NMC \cite{nmc} and BCDMS
\cite{bcdmsp,bcdmsd} collaborations. The preliminary results of a NLO
fit with fully correlated uncertainties \cite{qns} can be seen in
fig. \ref{f2nn} compared to other pdfs sets.
Our preliminary results appear to point
in the direction that 
the  uncertainties at small $x$ do not allow,
provided the current experimental data, to determine if
$q_{NS}(x,Q^2)$ grows at small $x$, as supported
by different theoretical arguments as well as 
by other global parton fits. However, more work is still needed
to confirm these results.
Only additional nonsinglet structure function data at small $x$
could settle in a definitive way this issue\footnote{
Like the experimental low $x$ deuteron structure function which
would be measured in an hypothetical electron-deuteron run
at HERA II, as it was pointed out during the workshop
by M. Klein and C. Gwenlan}.

Summarizing, we have described a general technique to parametrize
experimental data in an bias-free way with a faithful
estimation of their uncertainties, which has been successfully applied to
structure functions and that now is being implemented in the
context of  parton distribution. The next step will be to construct a full
set of parton distributions from all available hard-scattering data
using the strategy described  in this contribution.




%


\providecommand{\href}[2]{#2}\begingroup\raggedright\endgroup

\end{document}